\documentclass[twocolumn,aps,superscriptaddress,floatfix,final,showpacs,prl]{revtex4}
\usepackage{amsmath}
\usepackage{graphicx}
\usepackage{amssymb}
\usepackage{dcolumn}
\usepackage{bm}
\usepackage{amsfonts}
\usepackage{graphicx}
\usepackage{epsfig}
\usepackage{dcolumn}
\usepackage{bm}\begin{document}
\title{Universal quantum criticality at the Mott-Anderson transition}
\author{M. C. O. Aguiar}
\affiliation{Departamento de F\'{\i}sica, Universidade Federal de Minas Gerais, 
Avenida Ant\^onio Carlos, 6627, Belo Horizonte, MG, Brazil}
\author{V. Dobrosavljevi\'{c}}
\affiliation{Department of Physics and National High Magnetic Field 
Laboratory, Florida State University, Tallahassee, FL 32306, USA}


\begin{abstract}
We present a large N solution of a microscopic model describing the Mott-Anderson transition on a finite-coordination Bethe lattice. Our results demonstrate that strong spatial fluctuations, due to Anderson localization effects, dramatically modify the quantum critical behavior near disordered Mott transitions. The leading critical behavior of quasiparticle wavefunctions is shown to assume a universal form in the full range from weak to strong disorder, in contrast to disorder-driven non-Fermi liquid (``electronic Griffiths phase'') behavior, which is found only in the strongly correlated regime.  \end{abstract}
\maketitle

Both Anderson (disorder-driven) and Mott (interaction-driven) routes to localization play significant roles in materials close to the metal-insulator transition (MIT)~\cite{CIQPT}, but the full understanding of their interplay remains elusive. These effects are believed to be the cause of many puzzling features in systems ranging from doped semiconductors~\cite{nfl_sip_b} and two-dimensional electron gases~\cite{elihu_rmp}, to broad families of complex oxides~\cite{dagotto-science-2005}.

Early theories of the MIT~\cite{lr} focused on the stability of conventional metals to introducing weak disorder, but these Fermi-liquid approaches proved unable to describe strong correlation phenomena which took center stage in recent years. Indeed, the last two decades have seen significant advances in our understanding of ``Mottness'', not least because of the development of dynamical mean-field theory (DMFT)~\cite{dmftrev}, which has been very successful in quantitatively explaining many aspects of strong correlation. 

Because in its simplest form DMFT does not capture Anderson localization effects, several recent refinements were introduced, which proved capable of capturing both the Mott and the Anderson mechanisms. The conceptually simplest~\cite{tmt-book} such approach - the typical-medium theory (TMT) - provided the first self-consistent description of the Mott-Anderson transition, and offered some insight into its critical regime. For weak to moderate disorder, this theory found a transition closely resembling the clean Mott point, while only at stronger disorder Anderson localization modified the critical behavior. However, close scrutiny~\cite{ourtmt} revealed that some of these findings may be artifacts of the neglect of spatial fluctuations in this formulation. 

An alternative but technically more challenging approach to Anderson-Mott localization was dubbed ``Statistical DMFT'' ({\it stat}DMFT)~\cite{statdmft}. Here, only the strong correlations are treated in a self-consistent DMFT fashion, while disorder fluctuations are treated by a (numerically) exact computational scheme. While certainly much more reliable than TMT-DMFT, so far this method was utilized only in a handful of theoretical studies of the Mott-Anderson transition~\cite{hofstetter-2010prb,hofstetter-2011prb}, and the precise form of quantum criticality has never been explored in detail. 

In this letter, we present the first precise and detailed study of the quantum critical behavior of the Mott-Anderson transition in a Bethe lattice, within the framework of {\it stat}DMFT. We address the following physical questions, which we answered in a clear and reliable fashion: (1) Are there two distinct types of quantum criticality in this model, as TMT-DMFT suggested, or do the fluctuation effects restore universality within the critical regime? (2) How general is the disorder-driven non-Fermi liquid behavior (electronic Griffiths phase), and how does it relate to the relative strength of correlations and disorder? These results are obtained for a specific microscopic model, where our formulation proves exact in an appropriately defined large N limit. 

{\it Charge transfer model and {\it stat}DMFT} -- The charge transfer (CT) model has been used in the description of many multi-band systems near the Mott transition, including various oxides~\cite{ctm_oxides}, two-dimensional electron gases near Wigner-Mott crystallization~\cite{pankov-2008}, as well as doped  semiconductors~\cite{doped-book}. It consists of a two band model,  where one band represents weakly correlated conduction electrons and the other one is a narrow, strongly correlated band containing nearly Mott-localized electrons.  In the clean case, the Mott-insulating phase can be approached at odd integer filling by increasing the charge-transfer gap, which reduces the hybridization between the two bands. When the effective (correlation-renormalized) hybridization vanishes, the system undergoes a Mott metal-insulator transition.
\begin{figure}[t]
\begin{center}
\vspace{-12pt}
\includegraphics[scale=0.33]{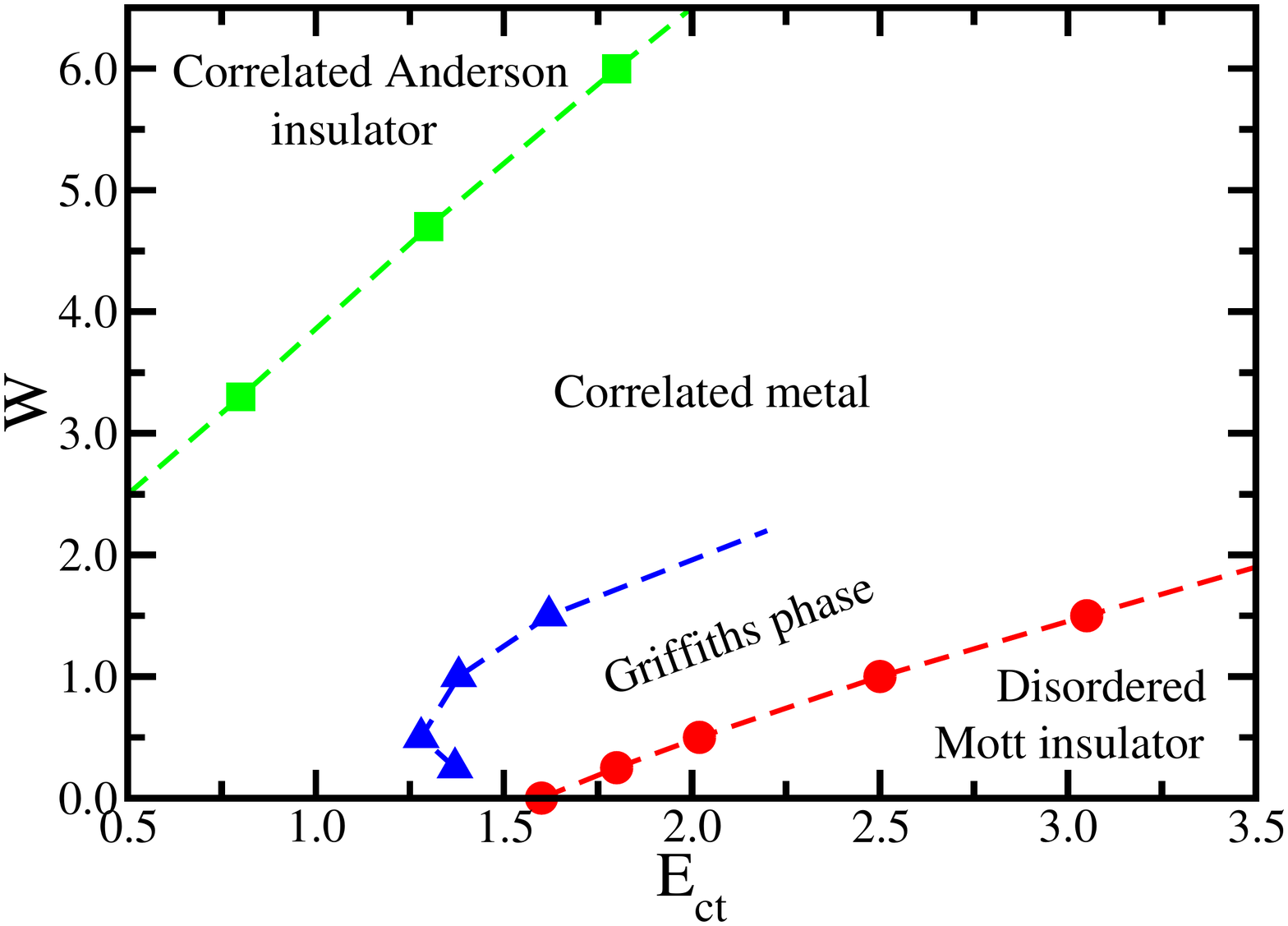}
\caption{(Color online) Phase diagram of the disordered CT model. $E_{ct}=-E_f$ is the
CT energy, which plays the role of Hubbard's $U$. The dashed lines are guides to the eyes.}
\label{fig1}
\vspace{-12pt}
\end{center}
\end{figure}

While many studies of the Mott-Anderson transitions concentrate on disordered single-band Hubbard models, we chose to focus on the CT model for two specific reasons. First, most real Mott systems are better described by the CT model than by the single-band Hubbard model, because it better incorporates the charge-transfer nature. Second, the simplest version of the CT model imposes the $U=\infty$ constraint (no double-occupancy) for the correlated band, a feature that considerably simplifies some technical aspects of our calculation, allowing a simple Gutzwiller-type variational solution of the model. This becomes formally exact in an appropriate large N limit, which is conveniently formulated within the standard slave-boson approach~\cite{coleman}.

The Hamiltonian of our disordered CT model corresponds to that of the 
disordered Anderson lattice model, which is given by
\begin{eqnarray}
H & = & \sum_{ij\sigma}\left[ (\varepsilon_i-\mu) \delta_{ij} -t  \right]
c^{\dagger}_{i\sigma}c_{j\sigma} +(E_f -\mu) \sum_{i\sigma} 
f^{\dagger}_{i \sigma} f_{i\sigma} \nonumber \\
  & + & V \sum_{i\sigma} 
\left( c^{\dagger}_{i\sigma} f_{i\sigma} + f^{\dagger}_{i\sigma}
c_{i\sigma} \right) +U \sum_i n_{fi\uparrow} n_{fi\downarrow}, \label{cth}
\end{eqnarray}
where $c_{i\sigma}^{\dagger}$ ($c_{i\sigma}$) creates (destroys) a conduction
electron with spin $\sigma$ on site $i$, $f_{i\sigma}^{\dagger}$ and $f_{i\sigma}$ 
are the corresponding creation and annihilation operators for a localized 
$f$-electron with spin $\sigma$ on site $i$, 
$n_{fi\sigma}=f_{i\sigma}^{\dagger}f_{i\sigma}$ is the number operator for
$f$-electrons, $t$ is the hopping amplitude, which is non-zero only for
nearest neighbors, $E_f$ is the $f$-electron energy, 
$U$ is the on-site repulsion between $f$-electrons, $V$ is the hybridization
between conduction and $f$-electrons, and $\mu$ is the chemical potential. 
Disorder is introduced through the on-site energies $\varepsilon_{i}$ for 
conduction electrons, which follow a distribution $P(\varepsilon)$, assumed to 
be Gaussian with zero mean and variance $W^2$.
Throughout this paper we use the half-bandwidth for 
conduction electrons as the unit of energy; the hybridization potential
is chosen to be $V=0.5$. 

Within the CT model, the average number of electrons per unit cell is equal to 
$1$, which can be enforced by adjusting the chemical potential and is
written as
\begin{equation}
\langle n_{ci} \rangle +\langle n_{fi} \rangle =1,
\end{equation}
where $n_{fi}=n_{fi\uparrow} + n_{fi\downarrow}$ gives the number of $f$-electrons
on site $i$, $n_{ci}=n_{ci\uparrow} + n_{ci\downarrow}$ is the corresponding 
number operator for conduction electrons, with $n_{ci\sigma}=c_{i\sigma}^{\dagger}
c_{i\sigma}$, and the averages are taken over the distribution 
$P(\varepsilon)$.

In this work we solve the above Hamiltonian using {\it stat}DMFT~\cite{statdmft}, 
in which the disordered lattice model is reduced to a self-consistent solution of an ensemble of single-impurity models located in the different sites $j$ of the lattice. The corresponding local effective actions take the form:
\begin{eqnarray}
& & S^{(i)}(j) \label{acaoef} \\ 
& & = \sum_{\sigma} \int_0^{\beta} d\tau
\int_0^{\beta} d\tau' f^{\dagger}_{j\sigma}(\tau) \left[\delta
(\tau-\tau') \left( \partial_\tau +E_f -\mu \right) \right.
\nonumber \\
& & + \left. \Delta_{fj} (\tau-\tau') \right] f_{j\sigma}
(\tau')+U \int_0^{\beta} d\tau n_{fj\uparrow} (\tau) n_{fj\downarrow} 
(\tau), \nonumber 
\end{eqnarray}
where the superscript $(i)$ indicates that site $i$ has been removed from
the lattice. The bath function, $\Delta_{fj} (\tau-\tau')$, is given by
\begin{equation}
\Delta_{fj}(i\omega)=\frac{V^2}{i\omega+\mu-\varepsilon_j-\Delta_{cj}(i\omega)},
\label{delta_result}
\end{equation}
where $ \Delta_{cj}(i\omega)=t^2\sum_{k=1}^{z-1}G^{(j)}_{ck}(i\omega)$,
$k \neq i$ in the summation, and $z$ is the coordination number 
(chosen to be $3$ in this work).
$G^{(j)}_{ck}(i\omega)$, the Green's function for conduction electrons, 
satisfies
\begin{equation}
G^{(j)}_{ck}(i\omega)=\frac{1}{i\omega+\mu-\varepsilon_k
-\Delta_{ck}(i\omega)-\Phi_k(i\omega)}, 
\label{condauto}
\end{equation}
where
\begin{equation}
\Phi_k(i\omega)=\frac{V^2}{i\omega+\mu-E_f-\Sigma_{fk}(i\omega)}
\label{phi}
\end{equation}
and $\Sigma_{fk}(i\omega)$ is the single-impurity self-energy, which is a
solution of an action similar to that given in eq.~(\ref{acaoef}), but for
site $k$ (instead of $j$). 

To solve the single-impurity problems of eq. (\ref{acaoef}), we use the slave-boson 
(SB) technique in the $U \rightarrow \infty$ limit~\cite{coleman}. In this 
case, the impurity Green's function can be written as 
\begin{eqnarray}
G_{fk}(i\omega)&=&\frac{Z_{k}}{i\omega-\varepsilon_{fk}-Z_{k}\Delta_{fk}(i\omega)},
\end{eqnarray}
where $Z_{k}$ is the local quasi-particle (QP) weight and $\varepsilon_{fk}$ is 
the renormalized $f$-electron energy. For technical details on solving the
{\it stat}DMFT equations for a closely related model see Ref.~\cite{localization03}. 

{\it Order parameter and phase diagram} -- 
To determine the phase diagram within $stat$DMFT, we examine the behavior of the {\it typical} value of the local density-of-states 
(LDOS) for conduction electrons at the Fermi energy ($\omega=0$). This order parameter~\cite{statdmft} measures the degree of localization of quasiparticle wavefunctions~\cite{tmt-book,ourtmt}.
It is given by
\begin{equation}
\rho_{typ}=\exp\{\langle \ln \rho_i(\omega=0) \rangle \},
\end{equation}
where $\langle ... \rangle$ denotes the (arithmetic) average over disorder and 
the most probable value is represented by the geometric average. Its vanishing directly indicates that wavefunctions are Anderson-localized.

\begin{figure}[t]
\begin{center}
\includegraphics[scale=0.29]{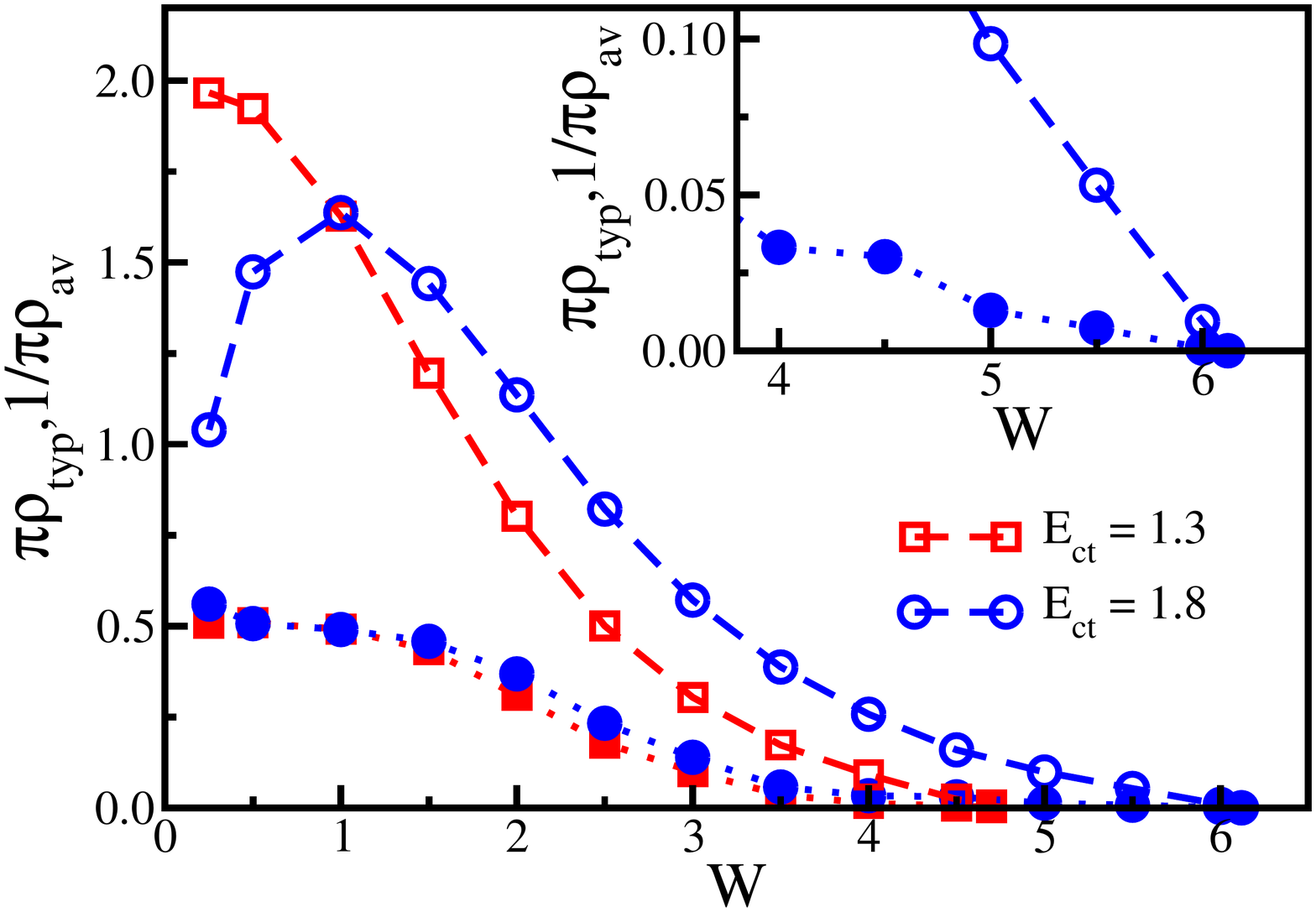}
\caption{(Color online) Typical (open symbols) and inverse of average (full 
symbols) values of 
the LDOS for conduction electrons close to the Fermi energy. Results are shown 
as a function of the disorder strength, $W$, for different values of the CT
energy, $E_{ct}$. Note that $E_{ct} = 1.8$ is larger than the critical $E_{ct}$
where the transition happens in the clean limit, which justifies the initial 
increase observed above for $\rho_{typ}$. The inset gives a zoom on the results 
close to the MIT.}
\label{fig2}
\end{center}\vspace{-16pt}
\end{figure}
Fig.~\ref{fig1} presents the phase diagram we obtained for the disordered 
CT model. Here, we show the phase boundary between the 
metallic and the Mott insulating phases, as well as that between the metal 
and the Anderson insulator; the region where a 
metallic Griffiths phase is seen is also shown, close to the Mott insulator. 
For our model, 
the charge-transfer gap, $E_{ct} = -E_f$, controls the effective strength of electron-electron interactions, i.e. plays the role of the Hubbard $U$.

In dramatic contrast to TMT-DMFT~\cite{ourtmt} and also
{\em stat}DMFT~\cite{hofstetter-2011prb} results for disordered Hubbard models, here (see Fig.~\ref{fig1}) we find a surprisingly broad intermediate metallic regime, spanning the region where correlations and disorder are comparable. This result may indicate that strong correlation effects have, for charge-transfer models, a more pronounced tendency to suppress Anderson localization than what was expected from Hubbard model studies. This finding is significant, because the CT model should be considered a more accurate representation of many real materials than the simple single-band Hubbard model. 

{\it Critical behavior near the disorder-driven transition} --
To obtain the phase boundary between the metallic and the Anderson insulating 
phases of Fig.~\ref{fig1}, we examined the behavior of $\rho_{typ}$ as a 
function of $W$, for different values of the CT energy; typical results 
are shown in Fig.~\ref{fig2}. For all values of $E_{ct}$ considered, 
$\rho_{typ}$ is found to decrease with disorder, reflecting Anderson 
localization effects. As the interaction parameter $E_{ct}$ increases, 
we observe that the MIT shifts to larger values of disorder. This behavior, 
which is caused by the tendency for disorder screening by correlation effects, is consistent with previous results 
obtained for the disordered Hubbard model within TMT-DMFT~\cite{ourtmt}. 
In addition, we find that the inverse of  $\rho_{av}=\langle \rho_i(\omega=0) \rangle$ (see Fig.~\ref{fig2}) vanishes at exactly the same $W$ value for which $\rho_{typ}$ vanishes, similarly as in {\em stat}DMFT studies of disorder-driven MIT in doped Mott insulators~\cite{statdmft}.
\begin{figure}[t]
\begin{center}
\includegraphics[scale=0.29]{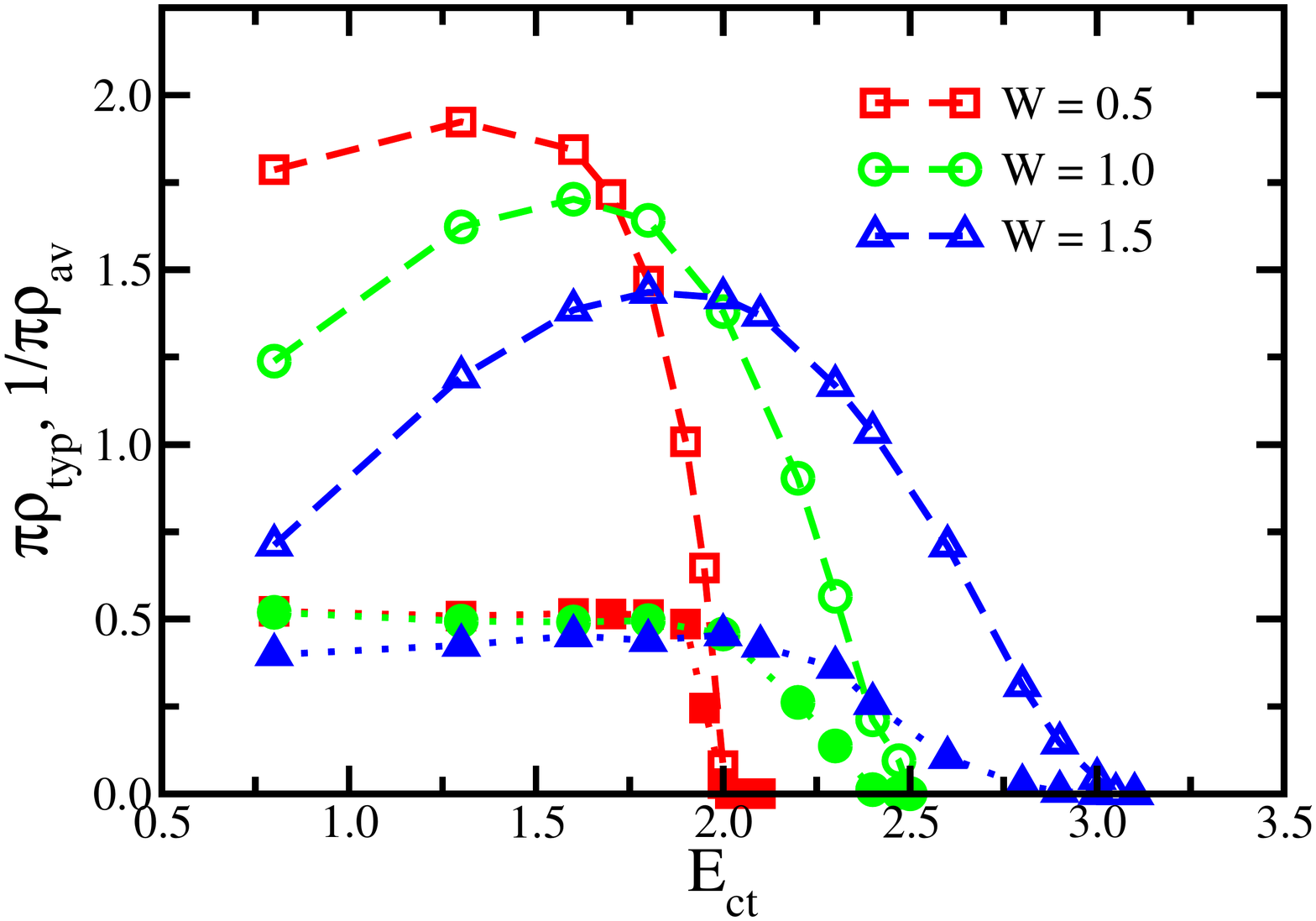}
\caption{(Color online) Typical (open symbols) and inverse of average (full 
symbols) values of 
the LDOS for conduction electrons close to the Fermi energy. Results are 
shown as a function of the CT energy, $E_{ct}$, for different values of the 
disorder strength $W$.}
\label{fig3}
\end{center}\vspace{-16pt}
\end{figure}

{\it Critical behavior near the Mott-like transition} --
The behavior of $\rho_{typ}$ as a function of $E_{ct}$ near interaction-driven (Mott-like) transitions is shown  in Fig.~\ref{fig3}. We find that, as disorder increases, the transition (given by  $\rho_{typ}=0$)  shifts to larger $E_{ct}$, consistent with previous results for disordered  Hubbard models~\cite{screening,ourtmt}, reflecting the tendency of disorder to broaden the Hubbard bands.  

The behavior of our localization order parameter  $\rho_{typ}$ shows very interesting behavior as the transition is approached. It initially increases with the interaction parameter $E_{ct}$, reflecting the correlation-induced screening of disorder within the metallic phase, an effect already found in previous studies~\cite{screening,ourtmt}. Very close to the transition, however, this tendency is {\em reversed}, and our order parameter  $\rho_{typ}$ starts to decrease and eventually vanishes, indicating localization of quasiparticle wavefunctions. Most remarkably, the critical behavior of both  $\rho_{typ}$ and the inverse of  $\rho_{av}=\langle \rho_i(\omega=0) \rangle$ vanish {\em linearly} at criticality - thus displaying precisely the same critical behavior as for the disorder-driven transition. This result is in dramatic contrast to the prediction of simpler DMFT or TMT-DMFT treatments~\cite{screening,ourtmt}, where $\rho_{typ}$ displays a finite jump at the disordered Mott transition. 

{\it The electronic Griffiths phase} --
The second order parameter identified by {\em stat}DMFT is the local quasiparticle (QP) weight $Z$~\cite{dmftrev,statdmft}; its vanishing indicates the destruction of the Fermi liquid by local moment formation (i.e. local Mott localization). In presence of disorder, this quantity displays spatial fluctuations, and must be characterized by an appropriate distribution function, $P(Z)$, which can assume  a power-law form $P(Z) \sim Z^{\alpha-1}$~\cite{statdmft,eric}. When the corresponding exponent $\alpha$ 
becomes smaller than $1$, this indicates~\cite{statdmft} disorder-driven non-Fermi liquid behavior~\cite{CIQPT}, dubbed the ``electronic Griffiths phase'' (EGP). 

We have carefully studied the behavior of $P(Z)$ across the phase diagram, and we find that EGP behavior is {\em not} found close to disorder-driven transitions, but is restricted to the vicinity of the interaction-driven (Mott-like) transition (see Fig. 1). Similar behavior has also been observed 
for the two-dimensional Hubbard model within {\it stat}DMFT~\cite{eric}. We also found that a finite fraction of sites display $Z=0$ within the Mott-like insulator, confirming the formation of localized magnetic moments (``Mott droplets'')~\cite{eric}. In contrast, all $Z$-s remain finite within the Anderson-like insulator, indicating the lack of local moment formation in this phase. 

{\it Comparison with DMFT results} -- For further insight in the role of Anderson localization in our model, we compare our {\em stat}DMFT results with those we obtained within conventional DMFT~\cite{screening}, which can be easily adapted for our model. Here, the average Green's function for conduction electrons satisfies  the following self-consistent equation
\begin{equation}
\bar{G}_c(i\omega)=\left< \frac{1}{i\omega+\mu-\varepsilon_k
-t^2 \bar{G}_c(i\omega)-\Phi_k(i\omega)} \right>.
\label{autodmft}
\end{equation}

\begin{figure}
\begin{center}
\includegraphics[scale=0.29]{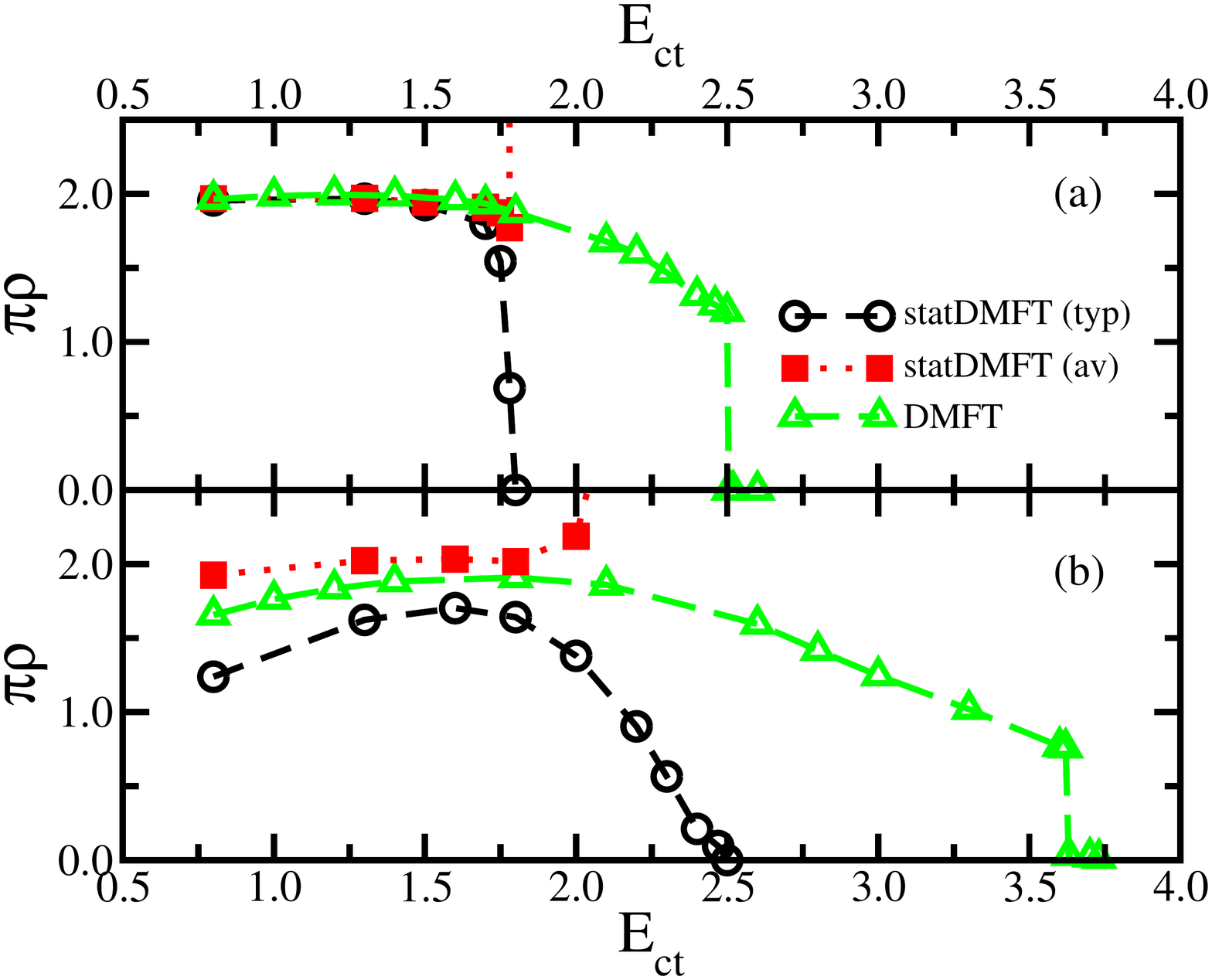}
\caption{(Color online) {\it Stat}DMFT typical ($\rho_{typ}$) and inverse of average
($1/\rho_{av}$) and DMFT average ($\rho_{dmft}$)
values of the LDOS for conduction electrons close to the Fermi
energy. Results are shown as a function of the CT energy, $E_{ct}$, for 
(a)~$W=0.25$ and (b) $W=1.0$. Note that both averages predicted by {\em stat}DMFT start to deviate from DMFT results only within the narrow critical region close to the Mott-like transition.}
\label{fig4}
\end{center}\vspace{-14pt}
\end{figure}

Since conventional DMFT is unable to capture Anderson localization effects but is able to describe Mott localization, the comparison is only meaningful and interesting close to the interaction-driven transition. Within DMFT the typical and the average values of LDOS coincide, and in 
Fig.~\ref{fig4} we compare the DMFT results with the corresponding quantities obtained from our {\em stat}DMFT calculation. Further from the transition, for small $E_{ct}$, results obtained from both theories essentially coincide, displaying correlation-induced screening. Closer to the transition, $\rho_{typ}$ obtained from  {\em stat}DMFT displays a maximum and starts to deviate from the DMFT result, and eventually vanishes much before DMFT would predict Mott localization (see Fig. 4). This clearly indicates that Anderson localization effects ``kick in'' only sufficiently close to Mott localization, and ultimately determine the resulting quantum critical behavior. We stress here that its precise (linear) form, though, is qualitatively different~\cite{statdmft} than in the non-interacting model, where  $\rho_{typ}$ vanishes exponentially for the Bethe lattice model we consider. 

We reiterate that these {\it disorder-induced} localization effects arise only within a narrow critical region close to the disordered Mott insulator. Its size can be estimated by examining how the location of $\rho_{typ}$ maximum evolves with disorder $W$. Remarkably, we find that the boundary of this critical region closely tracks the boundary of the EGP. This result makes perfect sense, as both Anderson localization effects and the EGP behavior reflect profound and dramatic spatial fluctuations of the relevant physical quantities. 

\noindent
{\it Conclusions} -- 
In summary, we carried out a detailed {\em stat}DMFT study of the $T=0$ critical behavior at the Mott-Anderson transition for a disordered charge-transfer model. Remarkably, identical critical behavior of the localization order parameter $\rho_{typ}$ is found in the close vicinity of both interaction-driven (Mott-like) and disorder driven (Anderson-like) transitions. This result demonstrates that (spatial) fluctuation effects captured by our {\em stat}DMFT, but not by simpler treatments, restore a degree of universality at this quantum critical point. In contrast, the thermodynamic behavior characterized by non-Fermi liquid features (electronic Griffiths phase) is found only near the Mott-like transition. 
Results obtained by experimental probes, such as Scanning Tunneling Microscope 
(STM), can be directly compared with our predictions for the behavior of the 
local DOS, and this is a fascinating direction for future experimental work.
Theoretically, finite temperature properties as well as the role of inter-site 
correlations remain to be examined in more detail, but these interesting 
questions are a challenge for future work.  

We thank Darko Tanaskovi{\' c} for useful discussions. We also acknowledge the 
HPC facility at Florida State University, where part 
of the results  were obtained. This work was
supported by CNPq and FAPEMIG (M.C.O.A.) and the NSF grant 
DMR-1005751 (V.D.).

\bibliographystyle{apsrev}

\bibliography{ct}

\end{document}